\documentclass[
 reprint,
 amsmath,amssymb,
 aps,
]{revtex4-2}

\usepackage{graphicx}
\usepackage{dcolumn}
\usepackage{bm}

\begin{document}


\title{Beyond Diffusion: A Causality-Preserving Model for Cosmic Ray Propagation}

\author{Norita Kawanaka}
\affiliation{National Astronomical Observatory of
Japan (NAOJ), 2-21-1, Osawa, Mitaka, Tokyo 181-8588,
Japan}
\affiliation{Department of Physics, Graduate School of Science Tokyo Metropolitan University 1-1,
Minami-Osawa, Hachioji-shi, Tokyo 192-0397}
\email{norita@tmu.ac.jp}
\author{Rohta Takahashi}%
\affiliation{%
National Institute of Technology, Tomakomai College, Tomakomai 059-1275, Japan
}%

\date{\today}

\begin{abstract}
We propose a brand-new formalism for the propagation of relativistic cosmic ray (CR) particles.  The propagation of CRs has often been described using the diffusion approximation, which has the drawback that the propagation speed of CRs near the source exceeds the speed of light.  By applying the analytic solution of the time-dependent distribution function of photons propagating while undergoing scattering, which we recently proposed, we have succeeded in formulating the propagation of relativistic CRs while preserving causality.  The obtained formulae give correct expressions both in the diffusion regime and the ballistic regime, as well as the transition between them.  They can be applied to the propagation of PeV CRs around their sources (PeVatrons), the propagation of ultra-high energy CRs, and the description of TeV gamma-ray halos around pulsars.
\end{abstract}

\maketitle


\section{Introduction}

The propagation of relativistic cosmic-ray (CR) particles in turbulent magnetic fields is often described by the diffusion equation.  In fact, CR particles experience multiple stochastic scattering by magnetic fields during their propagation in magnetized plasma, and from the theory of random walks, their transport can be approximated as the diffusion process with the diffusion coefficient of $D=(1/3)c\ell$, where $\ell$ is the scattering mean free path for a CR particle.  Since $\ell$ generally depends on the energy of a particle, the diffusion coefficient also depends on the energy, and its functional form is determined by the spectrum of turbulent magnetic fields.

However, the diffusion approximation cannot be applied to the scale shorter than $\ell$ because of the problem of superluminal propagation \cite{2007PhRvD..75d3001D, 2009ApJ...693.1275A, 2016ApJ...833..200L}.  Actually, in this regime the ballistic approximation is more relevant, where CR particles propagate coherently independent of their energy.  When discussing the spatial distribution of CRs in the vicinity of their source, one should take into account both the diffusion regime and the ballistic regime, as well as the transition between those two regimes.  The demand for such a formalism has been increasing particularly in recent years.  For example, thanks to the recent advent of the very high energy gamma-ray observations by Tibet AS$\gamma$ \cite{2019PhRvL.123e1101A, 2021PhRvL.126n1101A}, HAWC \cite{2020PhRvL.124b1102A, 2021NatAs...5..465A}, and LHAASO \cite{2023PhRvL.131o1001C, 2024ApJ...961L..43C, 2024ApJS..271...25C}, we are ready to search for PeVatrons, i.e., the sources of CRs with energy of $\gtrsim {\rm PeV}$, as well as determining their nature.  A PeVatrons can be identified via the observations of gamma-ray emission produced by hadronic interactions between CR particles and the ambient medium or inverse Compton scattering of ambient photons by CR electrons.  The energy spectrum and morphology of gamma-ray emission from a PeVatron will give us not only the information of injected CRs, but also that of CR propagation in the vicinity of the PeVatron.  As long as one adopt the diffusion approximation in describing the CR propagation, the problem of superluminal motion of high energy CRs is unavoidable.  Especially, since the mean free path length for PeV CRs is up to $30-300~{\rm pc}$, which corresponds to $0.2-2^{\circ}$ assuming the distance to the PeVatron as $8~{\rm kpc}$, a causality-preserving description for CR propagation is highly required in order to predict precisely the morphology of $\sim 100~{\rm TeV}$ gamma-rays originating from those PeV CRs with HAWC or LHAASO ($\sim 0.1-0.2^{\circ}$)\footnote{Note that the propagation of Galactic PeV CRs within the ballistic regime is not literally 'ballistic': they are moving along the magnetic field lines while spiraling around them.}.  The formalism beyond the diffusion approximation is also required to describe the propagation of ultra-high energy CRs (UHECRs) in the intergalactic space \cite{2006ApJ...643....8B, 2007ApJ...669..684B}, and may also be required to describe the TeV gamma-ray emissions around pulsar wind nebulae (``TeV halos'';\cite{2017Sci...358..911A, 2022NatAs...6..199L}), which are supposed to be created by CR electrons/positrons escaping PWNe via inverse Compton scattering \citep{2021PhRvD.104l3017R}.

There are several attempts to solve this problem by adopting the formalism other than the diffusion approximation.  One of them is to adopt the telegraph equation instead of the diffusion equation \cite{2013A&A...554A..59L, 2016RAA....16..162T}.  This formalism can limit the propagation speed below the speed of light by introducing an additional timescale that can separate the ballistic regime and the diffusion regime.  However, it has been claimed that the telegraph equation is not relevant for describing CR propagation because it does not conserve the total number of particles or their total energy \cite{2015ApJ...808..157M, 2016RAA....16..162T}.  Some studies adopted the J\"uttner function\cite{2007PhRvD..75d3001D, 2009ApJ...693.1275A, 2015PhRvD..92h3003P} , which is the phenomenological extension of the Maxwell distribution to the relativistic regime.  However, as had been proved in \cite{2007PhRvD..75d3001D}, this formalism has abandoned the Markov property to describe the particle propagation within a continuous diffusion model.  Moreover, their probability distribution function has not been confirmed to reproduce the particle distribution evaluated via Monte Calro simulations.  So far, no analytic formalism for the propagation of relativistic particles that can simultaneously describe the ballistic regime and the diffusion regime without violating the causality has been presented.  

In this work, we shall present the analytic formula to describe the propagation of relativistic CR particles in turbulent magnetic fields preserving the causality.  We also present the distribution functions of CR particles around the source injecting CRs continuously into the space, and compare them with those calculated from the diffusion approximation.  The case in which CRs lose energy during their propagation is also discussed.

\section{Formalism}
\subsection{Diffusion equation}
Before introducing our formalism, let us review briefly the diffusion equation describing the propagation of charged relativistic particles in turbulent magnetic fields.  We neglect the energy loss of particles and convection during the propagation.  Moreover, we consider the case that the diffusion coefficient $D$ is isotropic, is the function only of the energy of particles, $E$, and is independent of the position, $\bm{r}$ or time, $t$.  In this case, the CR diffusion equation can be written as
\begin{eqnarray}
\frac{\partial}{\partial t}f(t,E,\bm{r})=D(E)\nabla^2 f(t,E,\bm{r})+Q(t,E,\bm{r}),
\end{eqnarray}
where $f$ is the distribution function of CR particles and $Q$ is the source term, which represents the injection of CR particles from their source.  The diffusion coefficient $D$ and its energy dependence are determined by the strength of turbulent magnetic field and the spectrum of the turbulence.  Especially, the mean free path of particles scattered by turbulent magnetic fields, $\ell$, is the function of their energy and is related to the diffusion coefficient through $D(E)=c\ell(E)/3$.  Conventionally, the functional form of $D(E)$ in the interstellar medium is deduced from the observed CR spectra as
$D(E)\simeq D_0~{\rm cm}^2~{\rm s}^{-1}\left(E/\rm GeV \right)^{\delta}$, where $D_0\sim 10^{28}$ and $\delta\sim 0.3-0.6$ \cite{2016PhRvL.117w1102A}, with $\delta=1/3$ corresponding to a Kolmogorov-type turbulence, while $\delta=1/2$ corresponding to a Kraichnan-type turbulence \cite{2007ARNPS..57..285S, 2020PhR...872....1B}.  The Green function of this diffusion equation is given by $P_{\rm diff}(t_*,r_*)=\{3/(4\pi t_*)\}^{3/2}\exp\{-3r_*^2/(4t_*)\}$, where $t_*=ct/\ell$ and $r_*=\left|\bm{r}\right|/\ell$.

The diffusion approximation is, however, not valid when describing the propagation within the length scale shorter than the mean free path.  In the folllowing subsection we introduce a brand-new formalism without any approximation that can be applied to the CR propagation in any length scale.

\subsection{New Method}
We outline our new method for describing the propagation of CR particles as an alternative to the diffusion equation, and the resulting distribution function.  For details of the derivation of our formalism, we refer the readers to our paper \cite{2024ApJ...967L..10T}.

The collective behavior of relativistic particles such as CRs is described by the particle number density flux $N^\mu=(N, N^i)$ defined by 
\begin{equation}
    N^\mu=\int dN^\mu=\int \mathcal{F}(x^\mu, p^\mu)p^\mu dP
\end{equation}
where $x^\mu$ and $p^\mu$ are, respectively, the coordinate and the momentum of a particle, $\mathcal{F}$ is an invariant distribution function \citep[]{1966AnPhy..37..487L,1971agr.....2..331S,1971grc..conf....1E,1972grec.conf..201I}. The particle number in a volume element $dV$ at a time slice surface normal to a time-like unit vector $\hat{u}_\mu$ is given by the projection of $N^\mu$ onto the vector volume element $dV_\mu = dV \hat{u}_\mu$, i.e., $N(x^\mu)dV=N^\mu dV_\mu$. Since the time component of $N^\mu$ represents the particle number density $N$, dividing it by the total number of particles $N_{\rm all}$ on a constant time surface at $t=t_0$ gives the probability density function (PDF) $P(x^\mu)|_{t=t_0}$, which gives the probability that a particle exists in a certain spatial region at $t=t_0$, i.e. $P(x^\mu)|_{t=t_0}=N(x^\mu)/N_{\rm all}$. The PDF $P(x^\mu)$ is a function of the coordinate $x^\mu=(ct, \boldsymbol{r})$ of the particle. When there are particles to be scattered, it is convenient to introduce the coordinate $x^\mu_*$ normalized by the mean free path $\ell$ in the fluid rest frame, i.e., $x^\mu_*\equiv (t_*, \boldsymbol{r}_*)=(ct/\ell, \boldsymbol{r}/\ell)$. 

It has been reported that when a large number of relativistic particles are emitted into a static medium from a single point in space at a given instant and spread out in the medium with repeated isotropic elastic scattering, the PDF $P(t_*, \boldsymbol{r}_*)$ of the particles at a constant time surface can be written analytically \cite{2024ApJ...967L..10T}. In this case, the PDF has a spatially spherically symmetric distribution, i.e., the PDF is a function of $t_*$ and $r_*\equiv |\boldsymbol{r}_*|$. In this case, the PDF $P(t_*, r_*)$ of a time constant surface is described by the sum of the PDFs, $P_n(t_*, r_*)$, of particles that have experienced $n$ scatterings, i.e.,  
\begin{equation}
    P(x^\mu_*)=\sum_{n=0}^\infty P_n(x^\mu_*).
    \label{eq:Psum}
\end{equation}
The Fourier component of the PDF $P_n$ is described using the convolution theorem and is expressed with both a single scattering process and the initial conditions. Consequently, the PDF $P_n$ can be expressed in a variable-separated form, allowing the analytical solution to be obtained by solving the separated functions individually. 

Until recently, no analytical solution to the PDF has been found that describes the probability of existence in space-time of a large number of particles that repeatedly scatter and diffuse in 3-dimensional space \cite[e.g.,][]{2007JTP...20..769O,skogseid2011statistical,2024arXiv241019396L}. The PDF of a particle undergoing multiple scattering is known to obey the following equation (e.g., (7) in \cite{2024arXiv241019396L}),  
\begin{equation}
    P(x_*) = P_0(x_*) + \int p(x'_*) P(x_{*}-x'_*) dx'_{*},
    \label{eq:inteq}
\end{equation}
where $x_*=(t, \boldsymbol{r})$ is the 4-dimensional coordinate and $p(x_*)$ is the PDF of a particle with scattering-free ballistic motion, given by
\begin{equation}
    p(t_*, r_*)=\frac{e^{-r_*}}{4\pi r_*^2}\delta(t_*-r_*)\theta(t_*),\label{eq:pdfnoscatter}
\end{equation}
if the PDF is spherically symmetric. Furthermore, $P_0(x_*)$ is the PDF of a particle without scattering, and $P_0(x_*)=p(x_*)$. The equation (\ref{eq:inteq}) can be solved analytically in Fourier or Laplace-Fourier space by using the convolution theorem. Recently,  \cite{2024arXiv241019396L} obtain numerical solutions in real space by solving the inverse Fourier transform numerically. In other words, they find the numerical solutions of $P(x_*)$ in (\ref{eq:inteq}) \footnote{We have observed that when scattering is not very effective, numerical integration for (\ref{eq:inteq}) converges poorly and it is difficult to obtain highly accurate numerical solutions (e.g., double precision); we have also observed that analytical solutions obtained in Takahashi et al.(2024) \cite{2024ApJ...967L..10T} are faster to obtain than numerical solutions, which are more accurate.}. 

On the other hand, we point out that the PDF $P(x_*)$ given in Takahashi et al.(2024) \cite{2024ApJ...967L..10T} which gives an analytic solution of $P(x_*)$ in (\ref{eq:inteq}) as shown below. Since it is an analytical solution, it is possible to compute $P(x_*)$ exactly without some numerical errors. In our calculations, the PDF $P(x_*)$ is expressed in the expanded form by the scattering number $n$ as given in (\ref{eq:Psum}). The PDF $P_n(x_*)$ is calculated from the PDF $P_{n-1}(x_*)$ of particles experiencing $n-1$ times scattering by the following integral, 
\begin{eqnarray}
    P_n(x_{*}) 
    &=& \int p(x_{*}-x'_*) P_{n-1}(x'_*)dx'_{*} \nonumber\\ 
    &=& \int p(x'_*)P_{n-1}(x_{*}-x'_*) dx'_{*}.
\end{eqnarray}
Adding $n$ on both sides of this equation from $n=1$ to $\infty$, we obtain the following equation.
\begin{equation}
    \sum_{n=1}^\infty P_n(x_{*}) = \int p(x'_*) \sum_{n=1}^\infty P_{n-1}(x_{*}-x'_*) dx'_{*},
\end{equation}
where the LHS becomes $P(x_*)-P_0(x_*)$ and in the RHS we can calculate as 
\begin{equation}
    \sum_{n=1}^\infty P_{n-1}(x_*-x'_*)=\sum_{n=0}^\infty P_{n}(x_*-x'_*)=P(x_*-x'_*).
\end{equation}
Then, we obtain the equation in (\ref{eq:inteq}). This shows that the PDF $P(x_*)$ given in \cite{2024ApJ...967L..10T} is a solution of (\ref{eq:inteq}). The PDF $P_n(t_*,r_*)$ can be written analytically in variable separation form as follows,
\begin{equation}
P_n(t_*, r_*)=\frac{1}{4\pi}t_*^{n-3}e^{-t_*}V_n(r_*/t_*)\theta(t_*),
\end{equation}
where $\theta$ is a step function and $V_*(r_*/t_*)$ is an even function of $r_*/t_*$. In principle, the function $V_*(r_*/t_*)$ can be obtained analytically, but when $n$ is greater than about $n\gtrsim 10$, it becomes a very long analytical expression and cannot be handled practically. On the other hand, it is also possible to obtain an expression for the series expansion of the analytical expression, and to obtain an expression for the series expansion that achieves an arbitrary degree of accuracy (see, \cite{2024ApJ...967L..10T}).

\begin{figure}[t]
\includegraphics[width=0.95\columnwidth]{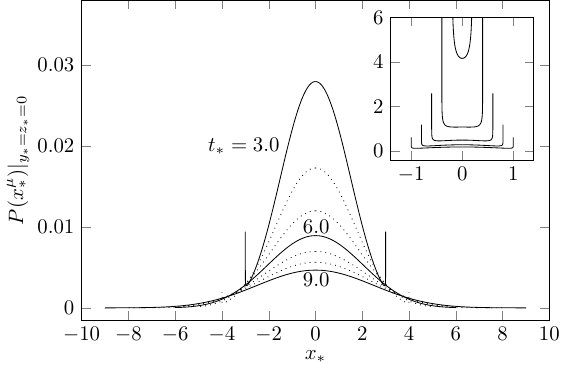}
\caption{\label{fig:PDF} The PDF $P(x^\mu_*)$ on the $x_*$ axis where $y_*=z_*=0$ of ultra-relativistic particles in a static medium at the time $t_*=0.2$, $0.4$, $0.6$, $0.8$, $1.0$, $3.0$, $6.0$ and $9.0$ (solid lines) and $t_*=4.0$, $5.0$, $7.0$ and $8.0$ (dotted lines). The inset shows the results for $t_*=0.2$, $0.4$, $0.6$, $0.8$, $1.0$ (from top to bottom). 
}
\end{figure}

Figure \ref{fig:PDF} shows the PDF of ultra-relativistic particles in a static medium from the time $t_*=0.2$ to $9.0$. Here, the velocity of the particle is approximated by the speed of light. For $t_*\lesssim 3.0$, the PDF describes the behavior of many particles when they are in ballistic motion at the speed of light, and the PDF shows spikelike peaks at $x_*=\pm t_*$, i.e., the location at $r_*=t_*$. These peaks represent the trace of ballistic particles, i.e., particles that do not undergo scattering. As time passes, most of  the particles are scattered at least once and the number of particles near $t_*=r_*$ gradually decreases. After the scattering is repeated, the PDF of the scattered particles changes to have a peak at $r_*=0$. Furthermore, as the particles are repeatedly scattered and the PDF evolves, the distribution of particles gradually approaches the PDF, $P_{\rm diff}(t_*, r_*)$, of particles in the diffusion limit. These PDFs are known to reproduce the results of Monte-Carlo simulations \cite{2024ApJ...967L..10T}, and have been confirmed to describe the state in which the particles propagate in a ballistic way, i.e., little or no scattering is occurring, and the state in which the diffusion approximation holds, as well as the transition between these states.

\begin{figure}[t]
\includegraphics[width=0.95\columnwidth]{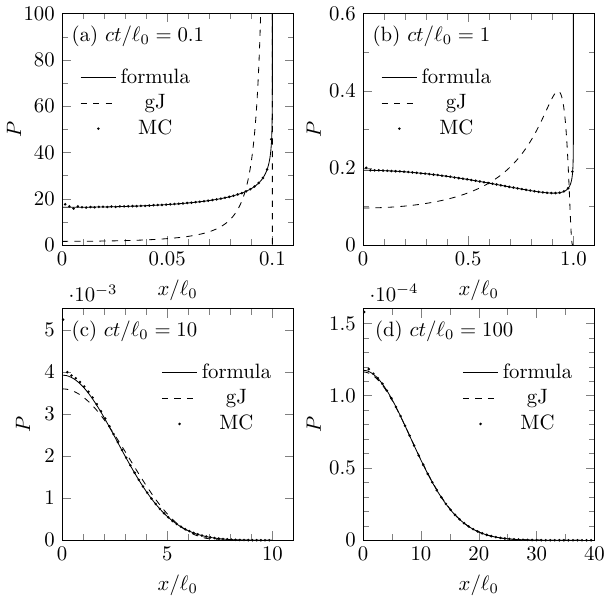}
\caption{\label{fig:compare} The PDF $P(x^\mu_*)$ (solid lines) and generalized J\"{u}ttner function given by the equation (24) in \cite{2009ApJ...693.1275A} (dashed lines) for times (a) $ct/\ell = 0.1$, (b) 1, (c) 10 and (d) 100. The results of relativistic Monte Carlo
simulations are shown (blank circles).}
\end{figure}

Figure \ref{fig:compare} depicts the PDFs of ultra-relativistic particles predicted from our formalism (solid lines) and the formalism introduced by \cite{2009ApJ...693.1275A} (based on the J\"{u}ttner function; dashed lines) for various times.  One can see that, especially at the times $ct/\ell = 0.1$ and $ct/\ell = 1$, there is a clear distinction between two predictions.  Moreover, one can see that the PDFs predicted from our formalism are consistent with the results of Monte Carlo simulations, which are also shown in the figures (blank circles).  Therefore, it is obvious that one should use our formalism to describe the CR propagation especially in the transition from the ballistic regime to the diffusion regime, instead of the formalism based on the J\"{u}ttner function.  In this study, we apply this PDF to the CR propagation in a medium and attempt a description of the continuous change from the ballistic state to the diffusion state of the particles. 

\section{Distribution of CR Particles around a source}

We consider a situation where ultra-relativistic particles are emitted in all directions from a single point with an emission rate $s_0(t_*)$ which represents the number of particles emitted per unit time at time $t_*$ in the particle source. After being emitted from the source, the particles spread spatially while repeatedly scattering through the medium around the source. Here, we assume an ultra-relativistic particle traveling at approximately the speed of light except when it is scattered. In such a setting, the number of particles $M(t_*, r_*)$\footnote{Here $M(t_*,r_*)=M(ct/\ell,r/\ell)$ corresponds to the distribution function of CR particles, $f(t,E,|\bm{r}|)$ with a fixed energy.} at time $t_*$ and distance $r_*$ from the CR source is given by
\begin{equation}
    M(t_*,r_*)=\int_0^{t_*-r_*}s_0(t_0)P(t_*-t_0,r)dt_0,
\end{equation}
where $t_0$ represents the time when the particle was emitted. Since $P(t_*, r_*)$ is expressed as a sum of $P_n(t_*, r_*)$, i.e., $P(t_*, r_*)=\sum_{n=0}^\infty P_n(t_*, r_*)$, $M(t_*, r_*)$ is also expressed as 
\begin{equation}
    M(t_*, r_*) = \sum_{n=0}^\infty M_n(t_*, r_*),
\end{equation}
where $n$ represents the number of scattering as denoted above and $M_n(t_*, r_*)=\int_0^{t_*-r_*}s_0(t_0)P_n(t_*-t_0,r)dt_0$. 

At time $t_*\ll 1$, $M(t_*,r_*)$ is approximated by $M_0(t_*,r_*)$. This is because for $t_*\ll 1$ most particles do not experience scattering and $P_0(t_*, r_*) \gg P_m(t_*, r_*) (m\ge 1)$. Here, $M_0(t_*, r_*)$ is calculated as 
\begin{eqnarray}
    M_0(t_*,r_*)
    &=&s_0(t_*-r_*)\frac{e^{-r_*}}{4\pi r_*^2}.
\end{eqnarray}
where we have used $P_0(t_*,r_*)=e^{-r_*}\delta(t_*-r_*)\theta(t_*)/(4\pi r_*^2)$. On the other hand, at time $t_*\gg 1$, many particles are experiencing scattering and are in a diffuse state. In this case, $M(t_*, r_*)$ is approximately calculated using the Green function in the diffusion limit $P_{\rm diff}$ as $M(t_*,r_*)=\int_0^{t_*-r_*}s_0(t_0)P_{\rm diff}(t_*-t_0,r_*)dt_0$. 

In this study, we consider two types of the particle sources. (i) The first is a steady source. In this case, we assume $s_0(t_*)=s_0\theta(t_*)$ where $s_0={\rm const}$. (ii) The second is a particle source that decays according to a power-law type function. In this case, we assume $s_0(t_*)=s_0 t_*^{-2}\theta(t_*)$ and this model simulates pulsar-like decay \citep[]{2010ApJ...710..958K}. In this model, to avoid singularity at $t_*=0$, assume $s_0=$const. for $t_*<t_0$. 

\begin{figure}[t]
\includegraphics[width=0.95\columnwidth]{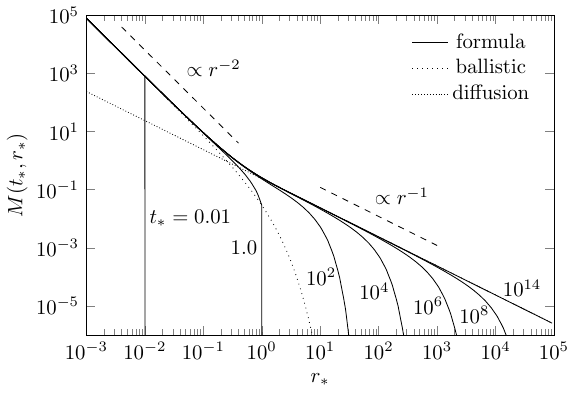}
\caption{\label{fig:Mstationary} The spatial distribution of ultra-relativistic particles emitted from type (i) source with $s_0=1$ from the time $t_*=0.01$ to $10^9$ (from left to right). 
}
\end{figure}

\subsection{Steady Source}
Let's consider first the case of a type (i) particle source. Figure \ref{fig:Mstationary} shows the time evolution of the spatial distribution $M(t_*, r_*)$ of particles in the type (i) source from the time $t_*=10^{-2}$ to $10^9$ (solid lines). In the present calculation, the distribution $M(t_*, r_*)$ is in the region $r_*\le t_*$ because the causality law is satisfied. At a region $r_*\ll 1$ (ballistic state), the spatial distribution of ultra-relativistic particles is approximated by 
\begin{equation}
    M_0^{(i)}(t_*,r_*)=s_0\frac{e^{-r_*}}{4\pi r_*^2}, 
\end{equation}
and the distribution in the region $r_*\lesssim 1$ (and $r_*\le t_*$) is proportional to $r_*^{-2}$. On the other hand, at time $t_*\gg 1$ (diffusion state), the spatial distribution of ultra-relativistic particles is approximated by 
\begin{eqnarray}
    M_{\rm diff}^{(i)}(t_*,r_*)
    &\approx& s_0\frac{3}{4\pi r_*}{\rm erfc}\left(\frac{\sqrt{3r_*}}{2}\right)~
    ({\rm for}~t_*\gg 1).
\end{eqnarray}
At time $t_*\gg 1$, this distribution forms a distribution proportional to $r_*^{-1}$ except at the edges of the distribution. The edges of the particle distribution $M(t_*, r_*)$ in the diffusion state spread spatially according to $r_*\propto \sqrt{t_*}$. In Figure \ref{fig:Mstationary}, we can see how the particle distribution gradually and continuously forms diffusion states. Eventually, a particle distribution $M(t_*, r_*)$ is formed that exhibits the ballistic state ($\propto r_*^{-2}$) near the source  ($r_*\lesssim 1$), the diffusion state ($\propto r_*^{-1}$) at $r_*>1$, and an intermediate state between them near $r_* \approx 1$. 

\begin{figure}[t]
\includegraphics[width=0.95\columnwidth]{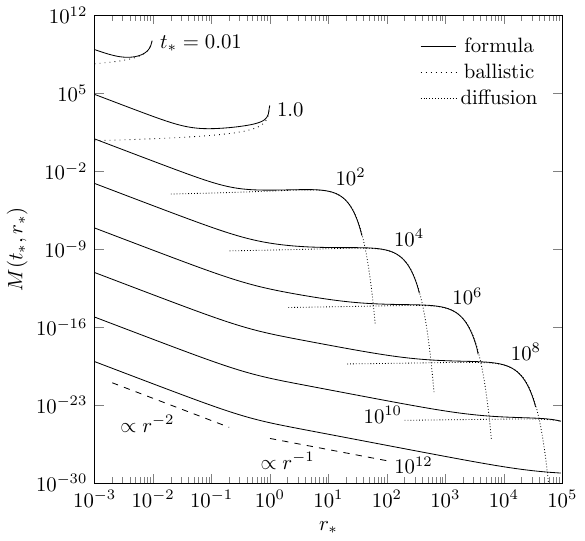}
\caption{\label{fig:Mpulsar} The spatial distribution of ultra-relativistic particles emitted from type (ii) source with $s_0=1$ and $t_0=0.001$ from the time $t_*=0.01$ to $10^{12}$ (from top to bottom). }
\end{figure}

\subsection{Pulsar-like Source}
Here, we consider the particle distribution $M(t_*, r_*)$ for a type (ii) source. In Figure \ref{fig:Mstationary}, we show the time evolution of the particle distribution $M(t_*, r_*)$ for the type (ii) source from the time $t_*=0.01$ to $10^{12}$ (solid lines). In this model, the particle emission rate decays according to $\propto t_*^{-2}$, so the spatial distribution of particles, $M(t_*, r_*)$, also roughly decreases according to $\propto t_*^{-2}$. At a region $r_*\ll 1$ (ballistic state), the particle distribution is approximated by 
\begin{equation}
    M_0^{(ii)}(t_*,r_*)
    =(t_*-r_*)^{-2}M_0^{(i)}(t_*,r_*),  
\end{equation}
and we can see the distribution which is proportional to $r_*^{-2}$ near the source. In the case of type (ii) source, the emission rate of the particles at a source decreases, so the amount of particles near the edges of the particle distribution $M(t_*, r_*)$ formed by particles emitted at earlier times is relatively larger than the amount of particles near the center. Thus, when $t_*\lesssim 1$, a spike-like peak at $r_*=t_*$, a ballistic signature, appears near the edge. This peak at $r_*=t_*$ has a shape similar to the distribution represented by $M_1(t_*, r_*)$ (dotted lines at $t_*=0.01$ and $1.0$ in Figure \ref{fig:Mstationary}), which is the distribution formed by particles that have experienced scattering only once. Therefore, when $t_*\lesssim 1$, the spatial distribution $M(t_*, r_*)$ is approximately represented by two regions; it is approximated by $M_0(t_*, r_*)$ near $r_*=0$ and by $M_1(t_*, r_*)$ near the edges of the distribution. On the other hand, when $t_*\gtrsim 10$, the distribution near the edge of the particle distribution is represented by $M_{\rm diff}^{(ii)}(t_*, r_*)=\int_0^{t_*-r_*}s_0(t_0)P_{\rm diff}(t_*-t_0,r)dt_0$, which corresponds to the diffusion state (densely dotted lines at time from $t_*=10^2$ to $10^{10}$ in Figure \ref{fig:Mstationary}). When $10\lesssim t_* \lesssim 10^5$, the spatial distribution is approximated by two regions; $M_0(t_*, r_*)$ near $r_*=0$ and $M_{\rm diff}(t_*, r_*)$ at the edge. As time elapses further and $t_*\gtrsim 10^6$, an intermediate region with a distribution proportional to $r_*^{-1}$ appears between the central region and the region of the edges. We have found that the spatial distribution at the intermediate region is approximated by 
\begin{equation}
    M_{\rm inter}^{(ii)}(t_*, r_*)=(t_*-r_*)^{-2}M_{\rm diff}^{(i)}(t_*, r_*),
\end{equation}
At this time, the spatial distribution consists of three regions; $M_0^{(ii)}(t_*, r_*)$ near $r_*=0$, $M_{\rm diff}^{(ii)}(t_*, r_*)$ at the edge, and $M_{\rm inter}^{(ii)}(t_*, r_*)$ at the intermediate region. 

From the above calculations, it can be seen that the shape of the spatial distribution $M(t_*, r_*)$ of particles around a CR source with type (i) and type (ii) emission rates $s_0(t_*)$ is different at any stage of the 
evolution. Thus, the shape of the spatial distribution $M(t_*, r_*)$ at a given instant can provide information on the time evolution of the emission rate $s_0(t_*)$ of particles at the source.

\section{Case with Energy Loss}

In this section we discuss the case with CR particles whose energy loss during propagation is not negligible.  Suppose that the energy loss of a CR particle during its propagation can be described as
\begin{eqnarray}
\frac{dE}{dt}=-b(E),
\end{eqnarray}
where $b(E)$ is the energy loss rate.  In this case, since the mean free path of a CR particle is no longer constant but varies with time, the PDF of CRs differs from the one obtained above. However, by introducing a new time coordinate $t_*$ and a new radial coordinate $r_*$ as defined below, the PDF can be derived in exactly the same manner as in Section II.B. 

In the presence of CR cooling, since the mean free path $\ell$ is generally a function of the CR energy $E$, it is useful to define the new time coordinate $t_*$ as 
\begin{eqnarray}
t_*
=\int_0^t \frac{cdt^{\prime}}{\ell(t^{\prime})}
=\int_{E}^{E_0}\frac{cdE^{\prime}}{\ell(E^{\prime})b(E^{\prime})},
\end{eqnarray}
and similarly, the new radial coordinate $r_*$ is defined as 
\begin{equation}
    r_*=\int_0^{r/c} \frac{cdt^{\prime}}{\ell(t^{\prime})}=\int_{E}^{E_0}\frac{cdE^{\prime}}{\ell(E^{\prime})b(E^{\prime})},
\end{equation}
where $E_0$ is the energy of a CR particle at the time $t^{\prime}=0$ whose energy is $E$ at the time  $t^{\prime}=t$ or $r/c$.  With these definitions, the PDF of a particles in ballistic motion without scattering, $p_*(t_*,r_*)$, can be described in the same way as $p(x_*)$ in the case without energy loss, Eq.(\ref{eq:pdfnoscatter}).  Using equations (3)-(7), one can obtain the PDF of particles in terms of $t_*$ and $r_*$, i.e., $P_*(t_*,r_*)$.

In order to obtain explicit relations between $t_*$ and $t$, and between $r_*$ and $r$, we assume the following parameterizations: 
\begin{eqnarray}
\ell(E)&=&\ell_0\left(\frac{E}{E_0}\right)^{\alpha}, \\
b(E)&=&b_0\left( \frac{E}{E_0} \right)^{\beta},
\end{eqnarray}
where $\ell_0$, $b_0$, $E_0$, $\alpha$, and $\beta$ are constants. For instance, when considering energy loss due to synchrotron radiation in the interstellar medium (ISM), with $D(E) \propto E^{1/3}$ as is often assumed, one may take $\alpha = 1/3$ and $\beta = 2$.

\begin{figure}[t]
\includegraphics[width=0.95\columnwidth]{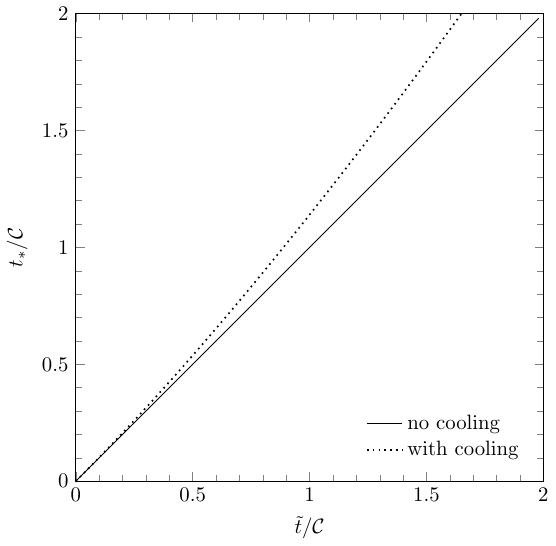}
\caption{\label{fig:tstt} 
The relationship between $t_*$ and $\tilde{t}$ is presented, where the solid line corresponds to the case without cooling ($p=1$), and the dotted lines represent cases with cooling ($p=4/3$). Both the vertical and horizontal axes, $t_*$ and $\tilde{t}$, are normalized by $\mathcal{C}$.
}
\end{figure}

If we define the normalized time $\tilde{t}$ and radial distance $\tilde{r}$ as
\begin{eqnarray}
\tilde{t}&=&\frac{ct}{\ell_0}, \\
\tilde{r}&=&\frac{r}{\ell_0},
\end{eqnarray}
then the transformed coordinates $t_*$ and $r_*$ can be expressed as 
\begin{eqnarray}
    t_*&=&\frac{\mathcal{C}}{p}\left[ \left(1+\tilde{t}/\mathcal{C}\right)^p-1\right], \\
    r_*&=&\frac{\mathcal{C}}{p}\left[ \left(1+\tilde{r}/\mathcal{C}\right)^p-1\right],
\end{eqnarray}
where
\begin{eqnarray}
    p&=&\frac{\alpha+\beta-1}{\beta-1}, \\
    \mathcal{C}&=&\frac{cE_0}{\ell_0 b_0(\beta-1)}.
\end{eqnarray}
Here, $p$ represents the effective power-law index, and $\mathcal{C}$ characterizes the dimensionless scale at which cooling becomes significant. Specific values of $\mathcal{C}$ for various particle energies will be evaluated later.

The coordinates $t_*$ and $r_*$ can be expanded in terms of $\tilde{t}/\mathcal{C}$ and $\tilde{r}/\mathcal{C}$, respectively, as 
\begin{eqnarray}
    t_*&=&\tilde{t}+\mathcal{C}(p-1)\bigg\{\frac{1}{2}(\tilde{t}/\mathcal{C})^2+\frac{1}{3!}(p-2)(\tilde{t}/\mathcal{C})^3
    \nonumber\\
    &&
    +O[(\tilde{t}/\mathcal{C})^4]\bigg\},\\
    r_*&=&\tilde{r}+\mathcal{C}(p-1)\bigg\{\frac{1}{2}(\tilde{r}/\mathcal{C})^2+\frac{1}{3!}(p-2)(\tilde{r}/\mathcal{C})^3
    \nonumber\\
    &&
    +O[(\tilde{r}/\mathcal{C})^4]\bigg\},
\end{eqnarray}
In the limit $\tilde{t}/\mathcal{C} \ll 1$ (or $\tilde{r}/\mathcal{C} \ll 1$), we recover $t_* \approx \tilde{t}$ (or $r_* \approx \tilde{r}$). In the case of $p=1$, which corresponds to the absence of energy loss, the relations reduce to $t_* = \tilde{t}$ and $r_* = \tilde{r}$. For synchrotron energy losses, $p = 4/3$. Figure~\ref{fig:tstt} shows the relationship between $t_*$ and $\tilde{t}$. It can be seen that $t_*$ deviates from $\tilde{t}$ when energy losses become significant. A similar relation holds between $r_*$ and $\tilde{r}$.

Let us evaluate the PDF in terms of physical coordinates, $\tilde{t}$ and $\tilde{r}$.  $P(\tilde{t},\tilde{r})$ and $P_*(t_*,r_*)$ should satisfy the following equation:
\begin{eqnarray}
    P(\tilde{t},\tilde{r})4\pi \tilde{r}^2 d\tilde{r}=P_*(t_*,r_*)4\pi r_*^2 dr_*.
\end{eqnarray}
Then, the PDF in the presence of cooling, $P(\tilde{t}, \tilde{r})$, is obtained as
\begin{equation}
    P(\tilde{t}, \tilde{r}) = P_*(t_*, r_*) \mathcal{F}
    \label{eq:PDFwithCooling}
\end{equation}
where $\mathcal{F}$ is the transformation coefficient arising from the coordinate restitution, given by
\begin{equation}
    \mathcal{F} \equiv \frac{r_*^2}{\tilde{r}^2} \frac{dr_*}{d\tilde{r}} = \frac{x^{p-1}(x^p-1)^2}{p^2(x-1)^2}
    =\frac{y^{1-1/p}(y-1)^2}{p^2(y^{1/p}-1)^2},
\end{equation}
where
\begin{eqnarray}
    x&\equiv&1+\tilde{r}/\mathcal{C},\\
    y&\equiv&1+pr_*/\mathcal{C}.
\end{eqnarray}
Equation (\ref{eq:PDFwithCooling}) allows us to calculate the PDF when the cooling effect is introduced.

\begin{figure*}[t]
\includegraphics[width=1.6\columnwidth]{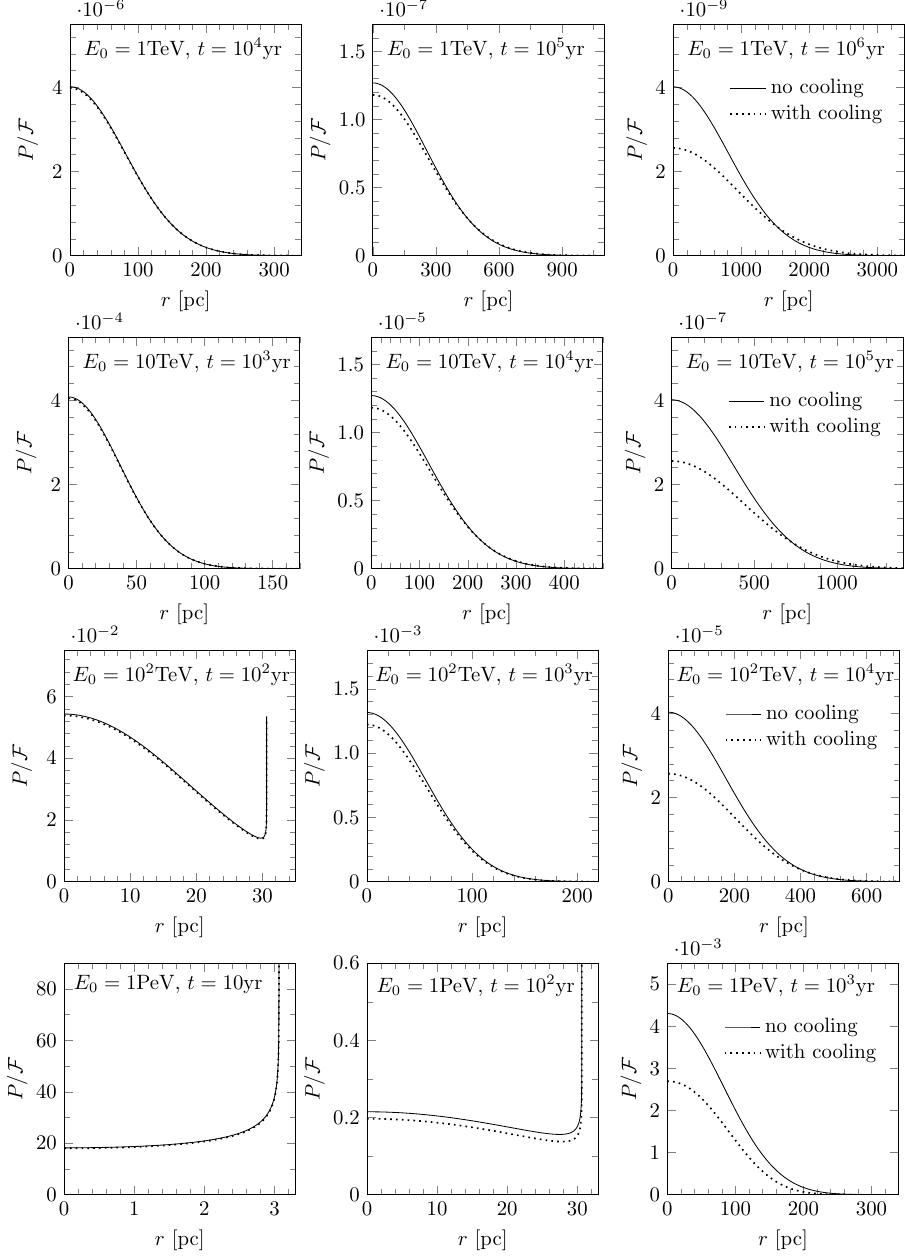}
\caption{\label{fig:cooling_PDF} 
The probability density function $P(\tilde{t}, \tilde{r})$, normalized by $\mathcal{F}$, for cases without cooling ($p = 1$, solid lines) and with cooling ($p = 4/3$, dotted lines). The calculations assume $E_0 = 1$ TeV, 10 TeV, 100TeV and 1 PeV (from top to bottom). Here, we assume $1{\rm yr}=31536000{\rm s}$. 
}
\end{figure*}

Specifically, to calculate the PDF when cooling effects are introduced, consider the following model that represents the energy dependence of the energy loss rate $b(E)$ and mean free path $\ell(E)$. When adopting the diffusion coefficient $D(E)=10^{28}\left( E/{\rm GeV} \right)^{1/3}~{\rm cm}^2~{\rm s}^{-1}$, one can describe the mean free path as a function of $E$ as 
$\ell(E)\simeq 10^{18}~{\rm cm} (E/{\rm GeV})^{1/3}$. On the other hand, the energy loss
rate $b(E)$ is the function of the electron/positron energy $E$, and in the Thomson limit it can be approximated as $b(E)=10^{-16}~{\rm GeV}~{\rm s}^{-1}\left(E/{\rm GeV} \right)^2$ \cite{1998PhRvD..59b3511B, 1998ApJ...493..694M}. 
Using these functional forms for $\ell(E)$ and $b(E)$, we obtain $p=4/3$, and $\mathcal{C}$ is evaluated for each energy scale of $E_0$ as follows,
\begin{eqnarray}
    \mathcal{C} \simeq 
    \left\{
        \begin{array}{ll}
        3.0\times 10^4  & ~~{\rm for}~~E_0=1~{\rm TeV} \\
        1.4\times 10^3  & ~~{\rm for}~~E_0=10~{\rm TeV} \\
        65             & ~~{\rm for}~~E_0=10^2~{\rm TeV} \\
        3              & ~~{\rm for}~~E_0=1~{\rm PeV} 
        \end{array}
    \right.
    , 
    \label{eq:lc}
\end{eqnarray}
which characterizes the timescale at which cooling becomes significant. 

Figure \ref{fig:cooling_PDF} compares the PDF with (dotted lines) and without (solid lines) cooling effects. We assume $E_0=1~{\rm TeV}$ (top), $10~{\rm TeV}$ (second from the top), $10^2~{\rm TeV}$ (third from the top), and $1~{\rm PeV}$ (bottom) as the initial particle energies at $t=0$. Figure~\ref{fig:cooling_PDF} demonstrates that, across all energy scales, the effects of cooling become significant after a timescale consistent with Equation (\ref{eq:lc}). For initial CR energies of 1 TeV, $10$ TeV, and $10^2$ TeV, the cooling effect becomes significant during the diffusion phase, whereas for 1 PeV, it starts to be significant in the intermediate regime between the ballistic and diffusion phases. In the presence of cooling, the mean free path gradually decreases, leading to more frequent particle scattering and a shorter scattering timescale. Consequently, the PDF evolves more rapidly than in the absence of cooling. The method proposed in this paper enables the analytical treatment of cooling effects for CR particles of arbitrary energy, across the ballistic, intermediate, and diffusion phases.

\section{Discussion and Summary}
We present a brand-new formalism to describe the propagation of relativistic CR particles in turbulent magnetic fields, which can deal with both the diffusion regime and the ballistic regime analytically as well as the transition between them, preserving the causality without any approximation.  We apply the analytical expressions for the PDF of photons that are multiply scattered in the uniform medium \cite{2024ApJ...967L..10T}, which are confirmed to be reproduced by Monte Carlo simulations.  As long as the energy of a CR particle does not change during its propagation, which is often a good approximation when it is a proton or a nucleus, its mean free path $\ell$ does not change either.  Therefore, the formulae for photons can be directly used in describing the propagation of CR protons or nuclei.  In addition, we also derive the PDF of CRs that lose their energies during propagation.  The results can be applied to the propagation of CR electrons and positrons that lose their energies via synchrotron emission and inverse Compton scattering, and to the propagation of UHECRs that lose their energy via collisional energy loss and adiabatic energy loss due to the Hubble expansion.  Applications of our formalism to these issues will be pursued in future works.

In this study, the time evolution of the spatial distribution $M(t_*, r_*)$ of relativistic CR particles around a particle source was investigated using PDF $P(t_*, r_*)$ of scattered particles.  There are some past studies that have proposed PDFs of CRs that avoid superluminal motion.  For example, \cite{2007PhRvD..75d3001D} proposed the PDF for particles in relativistic diffusion processes with non-Markovian properties making use of the J\"{u}ttner function. On the other hand, we consider that the PDF used in this study retains the causality and has Markovian properties because it satisfies the Chapman-Kolmogorov criterion \cite{2024ApJ...967L..10T}. Also, the PDF of the non-Markovian model of does not have peaks at the edges of the spatial distribution of particles even when the number of scattering is small (see, Figure 3 of \cite{2007PhRvD..75d3001D}), but the PDF we used has a distribution with spike-like peaks at $r_*=t_*$ for $t_*\lesssim 1$, as shown in Figure \ref{fig:Mstationary}.  Actually, these peaks also appear in the results of Monte Carlo simulations as shown in Fig. 2, which means that our formalism has a great advantage over the method using the J\"uttner function.

In the past studies, the observational data of the $\gamma$-ray halo around a pulsar is fitted by the diffusion model and/or the ballistic model \cite{2021PhRvD.104l3017R,2022ApJ...936..183B}. On the other hand, in our calculations, the ballistic and diffusion regimes are not described separately, and the PDF used in this study shows a gradual transition in the order of the ballistic regime, transition regime, diffusion regime. Since our PDF reproduces the results of relativistic Monte-Carlo calculations \cite{2024ApJ...967L..10T}, the transition from the ballistic regime to the diffusion regime is expected to progress gradually in the real situation as well.

There is also a previous study that describes the momentum distribution of particles during the transition from the ballistic state to the diffusion state by considering the first two moments of the Boltzmann equation  \cite{2015PhRvD..92h3003P}. Our PDF corresponds to the time component of the particle number density flux $N^\mu$ given by the momentum space integral of the invariant distribution function $\mathcal{F}(x^\mu, p^\mu)$ calculated on the basis of relativistic kinetic theory \cite{1966AnPhy..37..487L,1971agr.....2..331S,1971grc..conf....1E,1972grec.conf..201I}. Since some of co-authors in \cite{2024ApJ...967L..10T} have succeeded in deriving a solution of the invariant distribution function $\mathcal{F}(x^\mu, p^\mu)$ based on the analytic solution of the PDF of the scattering particles \cite{2024ApJ...967L..10T} and the solution of $\mathcal{F}(x^\mu, p^\mu)$ reproduces the results of the Monte-Carlo simulations (private communications with co-authors of \cite{2024ApJ...967L..10T}), we expect that our solution is consistent with the solution derived from the relativistic Boltzmann equation.  Moreover, the solution $\mathcal{F}(x^\mu, p^\mu)$ is useful when calculating the photon emissions associated with CRs because the angular distribution of CRs is anisotropic in the ballistic regime, which makes the angular distribution of photons produced via hadronic interaction or inverse Comption scattering.  Therefore, the expression of $\mathcal{F}(x^\mu, p^\mu)$ enables us to predict the gamma-ray spectra and morphology around CR sources, including PeVatrons and pulsar wind nebulae.  

This work is supported by JSPS KAKENHI grant Nos.  16K05302,  19H00697, 21H01132 (R.T.) and 22K03686 (N.K.).

\bibliography{apssamp}

\end{document}